\documentclass[5p, onecolumn]{elsarticle}

\journal{Journal of Systems and Software}
\usepackage{natbib}
\usepackage{amsmath}
\usepackage[hyphens]{url}
\usepackage{graphicx}
\usepackage{array}
\usepackage{multirow}
\usepackage{verbatim}
\usepackage{hyphenat}
\usepackage{float}
\usepackage{tcolorbox}
\usepackage{enumitem}
\usepackage{balance}
\usepackage{todonotes}
\usepackage{fontawesome}
\usepackage{threeparttable}
\usepackage{listings}
\usepackage{booktabs}
\usepackage{color, colortbl}
\usepackage{marvosym}

\newcommand\SelectedTools{84~}
\newcommand\SelectedPS{254~}
\newcommand\SelectedGray{203~}
\newcommand\SelectedWhite{51~}

\newcommand\SnowballingWhite{26~}
\newcommand\SnowballingGray{4~}

\newcolumntype{c}[1]{>{\centering\arraybackslash}p{#1}}

\usepackage{todonotes}
\usepackage{footmisc}

\usepackage{color}

\definecolor{gray50}{gray}{.5}
\definecolor{gray40}{gray}{.6}
\definecolor{gray30}{gray}{.7} 
\definecolor{gray20}{gray}{.8}
\definecolor{gray10}{gray}{.9}
\definecolor{gray05}{gray}{.95}

\newlength\Linewidth
\def\findlength{\setlength\Linewidth\linewidth
	\addtolength\Linewidth{-4\fboxrule}
	\addtolength\Linewidth{-3\fboxsep}
}
\newenvironment{rqbox}{\par\begingroup
	\setlength{\fboxsep}{5pt}\findlength
	\setbox0=\vbox\bgroup\noindent
	\hsize=0.95\linewidth
	\begin{minipage}{0.95\linewidth}\normalsize}
	{\end{minipage}\egroup
	\textcolor{gray20}{\fboxsep1.5pt\fbox
		{\fboxsep5pt\colorbox{gray10}{\normalcolor\box0}}}
	\endgroup\par\noindent
	\normalcolor\ignorespacesafterend}

\newcounter{Finding}
\stepcounter{Finding}

\newcommand{\quickwordcount}{%
  \immediate\write18{texcount -1 -sum -merge \jobname.tex > \jobname-words.sum }%
  \input{\jobname-words.sum} words%
}

\begin{document}
\begin{frontmatter}

\title{Toward End-to-End MLOps Tools Map: A  Preliminary Study based on a Multivocal Literature Review}

\author[TUNI]{Sergio Moreschini}
\ead{sergio.moreschini@tuni.fi}
\author[UNISA,TUNI]{Gilberto Recupito}
\ead{grecupito@unisa.it}
\author[OULU]{Valentina Lenarduzzi}
\ead{valentina.lenarduzzi@oulu.fi}
\author[UNISA]{Fabio Palomba}
\ead{fpalomba@unisa.it}
\author[TUNI]{David Hästbacka}
\ead{david.hastbacka@tuni.fi}
\author[OULU,TUNI]{Davide Taibi}
\ead{davide.taibi@oulu.fi}

\address[TUNI]{Tampere University, Finland}
\address[UNISA]{Software Engineering (SeSa) Lab, Department of Computer Science, University of Salerno, Italy}
\address[OULU]{University of Oulu, Finland}




\begin{abstract}
MLOps tools enable continuous development of machine learning, following the DevOps process. Different  MLOps tools have been presented on the market, however, such a number of tools often create confusion on the most appropriate tool to be used in each DevOps phase. To overcome this issue, we conducted a multivocal literature review  mapping \SelectedTools MLOps tools identified from \SelectedPS Primary Studies, on the DevOps phases, highlighting their purpose, and possible incompatibilities. The result of this work will be helpful to both practitioners and researchers, as a starting point for future investigations on MLOps tools, pipelines,  and processes. 

\end{abstract}


\end{frontmatter}

\section{Introduction}
\label{sec:intro}

MLOps (Machine Learning Operations) is a practice that combines the best practices of software engineering and data science to manage the end-to-end lifecycle of machine learning models, integrating them into traditional software. MLOps involves the use of tools and techniques to automate the building, testing, deployment, and monitoring of machine learning models. The goal of MLOps is to improve the speed, quality, and reliability of machine learning models while reducing the risk of errors. 

The use of MLOps tools is becoming increasingly important as the number of machine learning models being deployed in production environments continues to grow. These tools help data scientists and engineers to streamline the process of building, testing, and deploying machine learning models, and make it easier to monitor and maintain models in production. This allows organizations to be more agile in their use of machine learning, and to quickly adapt to changing business requirements and feedback from users. 

Different MLOps tools have been introduced in the market~\cite{Recupito2022}, and a large number of practitioner's conference started to introduce MLOps and related tools.

However, the very large number of MLOps tools, might introduce issues to organizations that need to select a subset of them to create their pipelines. Moreover, not all the tools can be combined together, introducing other problems during the adoption of them. 

In our previous work~\cite{Recupito2022}, we proposed a multivocal literature review to investigate the characteristics of MLOps tools mentioned by the first 102 hits of Google Search, and by Google Scholar. However, we pointed out our previous work had several limitations, including a limited scope of the search, and the lack of snowballing on the selected sources. 

The ultimate goal of the study is to provide researchers and practitioners with some preliminary insights into the tools available in the context of an end-to-end machine learning pipeline. On the one hand, the results of the study might be beneficial for researchers interested in understanding the current support provided to machine learning engineers in the wild. On the other hand, our findings might pose the basis for further studies aiming at eliciting the practitioner's perspective on where they would use the available tools in practice and how they would create an ideal pipeline to more effectively support end-to-end machine learning. 


Therefore, we conduct an internal differentiated replication~\cite{CarverReplication2014} of our previous work, overcoming the limitations, and extending it towards the identification of the different DevOps phases where each tool can be applied, identifying both compatibilities and incompatibilities. The main contribution of this work is a Graphical DevOps map of MLOps tools, useful to practitioners and researchers to easily identify useful tools for each DevOps phase. 

The remainder of this paper is structured as follows. 
Section~\ref{sec:Background} introduced the background on DevOps and MLOps. Section~\ref{sec:Original} describe the replicated study~\cite{Recupito2022} and highlights the differences with our work. Section~\ref{sec:study} presents the research questions targeted by our study and the methods adopted to address them. The results are presented in Section~\ref{sec:Results} and further elaborated in Section~\ref{sec:Discussion}. Section~\ref{sec:Threats} identifies and reports the threats to the validity of our study. Section~\ref{sec:RW} discusses the related work in the field of MLOps, highlighting how our work advances the state of the art. Finally, Section~\ref{sec:Conclusion} draws conclusions and highlights future works.

\section{Background}
\label{sec:Background}

DevOps is a set of practices and tools that aim to improve collaboration and communication between development and operations teams in order to accelerate the software delivery process. The goal of DevOps is to increase the speed and quality of software releases while also reducing the risk of failures and downtime~\cite{bass2015devops}.

The DevOps process typically includes the following phases:


\begin{itemize}
    \item \textbf{Plan.} All the activities conducted before Code development. Identification of the business requirement and collection of end-user feedback. The goal is to create a product roadmap for future development.
    \item \textbf{Code.} All the activities related strictly to the Code development phase. The tools used at this stage include plugins to aid the development process.
    \item \textbf{Build.} Once a coding task has been completed, a developer commits the code to a shared code repository to produce a build.
    \item \textbf{Test.} The build is deployed in a test environment to undergo several types of testing. Such tests include acceptance tests, security tests, integration tests, performance tests, etc.
    \item \textbf{Release.} After a build has successfully passed all the automated and manual tests, it is ready for deployment in the production environment and the operations team can schedule the releases.
    \item \textbf{Deploy.} The build is ready and it is released into production. A set of tools are used to  automate the release process.
    \item \textbf{Operate.} At this stage the release is out and already used by customers. The operations team is responsible for server configuring and provisioning.
    \item \textbf{Monitor.} The whole pipeline is monitored based on the data acquired at the Operate stage.  
\end{itemize}

Machine Learning Operations (MLOps) is a set of practices and tools that aim to improve the collaboration and communication between data scientists and IT/DevOps teams when building, testing, and deploying machine learning models. It is designed to help organizations manage the complexity of deploying machine learning models in production environments and ensure that the models are reliable, scalable, and maintainable.

One key difference between MLOps and traditional DevOps is that MLOps focuses specifically on the needs of machine learning models, which can be very different from traditional software. For example, machine learning models require much more data to be trained and are often more computationally intensive than traditional software. Additionally, machine learning models are also subject to different types of errors, such as overfitting and bias, which can be difficult to detect and fix.

Another key difference is that MLOps often involves more complex, data-intensive workflows. Data scientists may need to work with large amounts of data, perform feature engineering, and experiment with different model architectures, all of which can be time-consuming and resource-intensive. MLOps helps bridge the gap between data scientists and IT/DevOps teams by automating many of the tasks involved in building, testing, and deploying machine learning models, such as data preprocessing, model training, and model serving.

MLOps also helps with model governance and management. Machine learning models need to be deployed, monitored, and maintained, and MLOps provides the necessary infrastructure and tooling to automate these tasks. It also helps keep track of the models' versions, and performance metrics, and provides transparency of the data used to build the models.

In summary, MLOps is a set of practices and tools that help organizations manage the complexity of deploying machine learning models in production environments. It focuses on the specific needs of machine learning models and helps bridge the gap between data scientists and IT/DevOps teams by automating many of the tasks involved in building, testing, and deploying machine learning models. It also helps with model governance and management, providing the necessary infrastructure and tooling to automate these tasks.

\begin{figure}[ht]
    \centering
    \includegraphics[width=.40\textwidth]{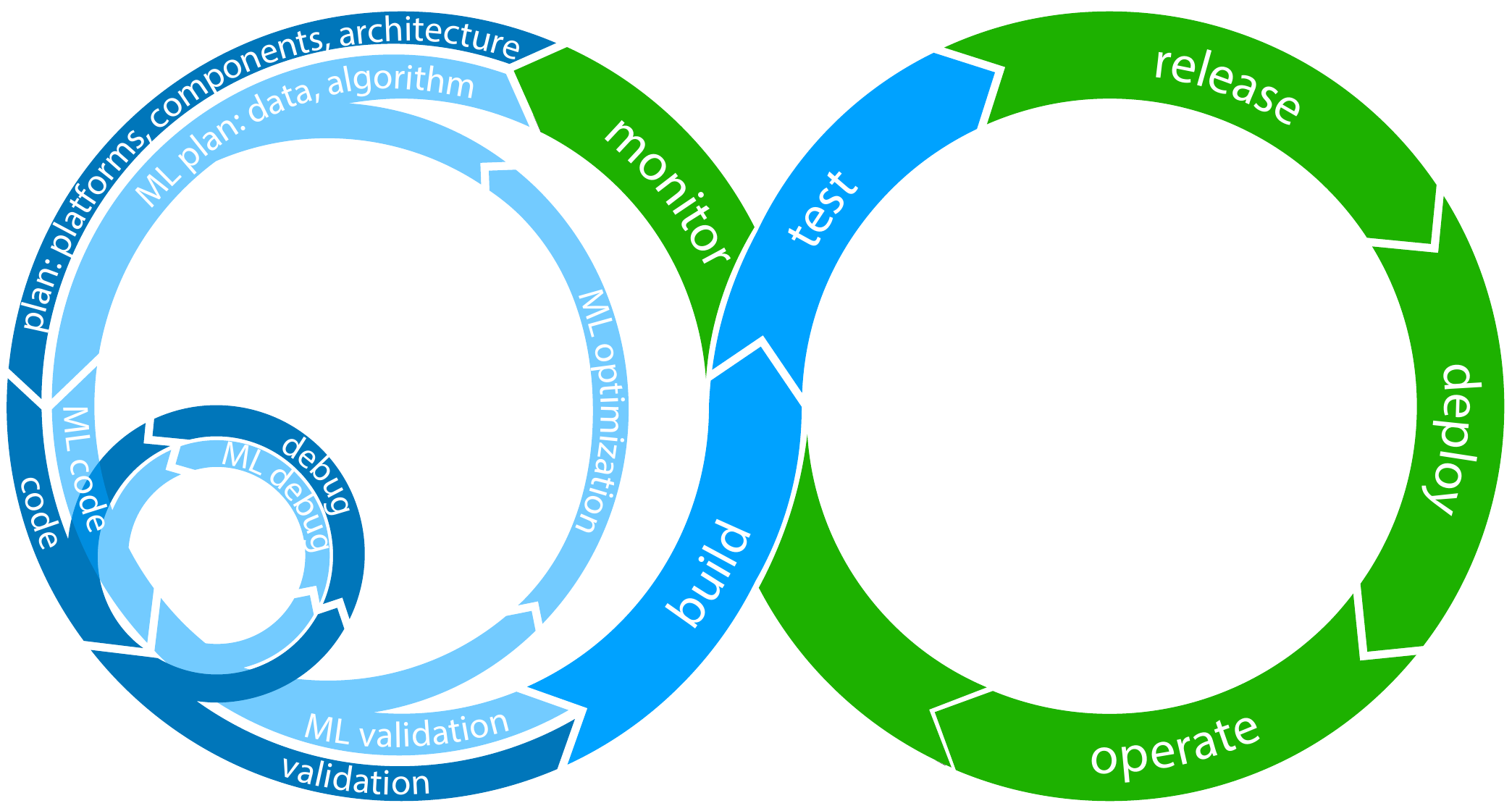}
    \caption{MLOps infinite loop \cite{Moreschini22}}
    \label{fig:MLOps}
\end{figure}

A graphical representation of the MLOps process, proposed by Moreschini et al.~\cite{Moreschini22} is illustrated in (Figure~\ref{fig:MLOps}). 

\section{The Replicated Study}
\label{sec:Original}
\begin{table*}[h!]
\caption{Studies Comparison}
\label{tab:DesignComparison}
\footnotesize
\centering
\begin{tabular}{p{4cm}|p{6.3cm}|p{6.3cm}} \hline 
& \textbf{Recupito et al.}~\cite{Recupito2022} & \textbf{Our Study} \\ \hline 
RQs & - List of MLOps tools & - Mapping of MLOps tools into the MLOps phases \\
& - Main features of MLOps tools & \\ \hline 
Search strings & \multicolumn{2}{l}{We added more search terms to the search string including ''ML ops'', and ''ML-ops''} \\ \hline
Bibliographic references & Google Scholar  &  Google Scholar, ACM digital Library, IEEEXplore Digital Library, Science Direct, Scopus, and Springer link\\ \hline 
Gray literature search engines & Google & Google, Twitter, Reddit, and Medium\\ \hline 
Snowballing & Not performed & Performed \\ \hline
\multirow{8}{*}{Inclusion criteria (PS)} & - The study discusses components of minimal end-to-end MLOps workflow(s) & - PS mentioning tools that can
be used to create or compose MLOps pipelines\\
& - The study discusses the practice of MLOps or ML-based applications& \\
& - The study refers to the implementation of MLOps tool(s) & \\
& - The study discusses experience, opinions, or practices on
MLOps pipeline(s)& \\ \hline 
\multirow{9}{*}{Exclusion criteria (PS)} & - The study does not offer details concerning the design or implementation of MLOps tool(s) & - Research Plans, roadmaps, vision papers, Master Thesis, Books  \\
& - The study solely offers the design of a specific component of ML pipeline(s) & - Less than 750 followers for the author (only for medium - gray literature) \\
& - The study does not offer or refer to details concerning ML automation &  \\
& - The study refers to commercial platform(s) that offer(s) MLOps applications to sell their services for development and deployment &  \\ \hline 
Inclusion criteria (Tools) & Data extraction to model monitoring tools & End-to-end MLOps tools, including all the phases of the MLOps process (Figure~\ref{fig:MLOps}) \\ \hline 
Exclusion Criteria (tools) & 1) The MLOps tool does not provide the documentation or does not list characteristics & - Not essentials (Traning Orchestration, Explainable AI, FeatureStore) \\
 & - The MLOps tool is mentioned less than five times by the selected studies & - Tools breaking the automatization (data generation, data labelling, data monitoring) \\ 
 & & -  Addons \\
 & & -  Not reliable (ex. not enough stars) \\ \hline 
\multirow{2}{*}{Tools classification} & - main MLOps characteristics & - MLOps pipeline steps covered by each tool \\ 
& - data management features related to the ML models & - Compatibilities and incompatibilities between tools \\
& - ML model management features & - Possible combinations of tools for MLOps pipelines\\ \hline
Selected primary studies & 60 (6 white and 54 gray) & \SelectedPS  (\SelectedWhite white and \SelectedGray gray)\\ \hline
MLOps Tools & 13 & \SelectedTools\\\hline 

\end{tabular}
\end{table*}
In this Section, we summarize the replicated study~\cite{Recupito2022} and the reasons why we replicated it, highlighting the differences with our work. 

We decided to replicate the study~\cite{Recupito2022},  since – as far as we know – this is the only secondary study that identified a set of MLOps tools and their characteristics. 

The replication is performed following the guidelines proposed by Carver et al.~\cite{CarverReplication2014}.

Recupito et al.~\cite{Recupito2022} performed a Multivocal Literature Review investigating the MLOps tools that can be adopted in a ML pipeline from the data extraction to the model monitoring phases. They considered gray and white literature sources. They provided the tools list and their main distinctive features in general features, focusing on 1) main MLOps characteristics, 2) data management features related to the ML models, and 3) ML model management features. Moreover, they classified the tools into cloud-based ML platforms, orchestration platforms, or TensorFlow extended.
They extracted white literature from Google Scholar  and the gray literature considering the first 102 hits of Google Search. From these 102 hits, they retrieved 96 websites including blog posts and developers' websites, 9 GitHub repositories, and 5 YouTube videos. After applying inclusion and exclusion criteria, they included a total of 60 sources, of which 42 were websites/blogs, 7 GitHub Projects, 5 YouTube videos, and 6 peer-reviewed sources from Google Scholar.

As a result, they extracted 13 tools from the 60 selected sources.

\subsection{Differences with this  study}
Differently from our original study~\cite{Recupito2022}, we considered tools that can be adopted in all the steps of the MLOps pipeline, we extended the search strings and adopted six bibliographic sources for the white literature instead of one, and three search engines for the gray literature instead of one. 

Moreover, we performed the snowballing process, and we relaxed inclusion and exclusion criteria to obtain as many relevant papers as possible. 
As a result, we included \SelectedPS sources from which to extract tools in comparison to 60 considered by~\cite{Recupito2022}, and obtained a list of \SelectedTools tools instead of 13. 

Besides the larger amount of tools we classified, the most important result lies in the different focus of the replicated study. While in our previous work~\cite{Recupito2022} we identified the main features of the tools, we focused on the MLOps phases where the tools can be used, and on their compatibilities and incompatibilities when creating tools pipelines. 

The complete comparison between this work and the replicated one~\cite{Recupito2022} is reported in Table~\ref{tab:DesignComparison}.

\section{Study Design}
\label{sec:study}
The goal of our study is to map all the MLOps tools with the purpose of outlining a fully-supported MLOps pipeline. 
Therefore, in order to investigate the aforementioned goal, we formulated the following Research Question (RQ):

\vspace{2mm}
\begin{tabular}{p{0.5cm}p{7.2cm}}
\textbf{RQ}  & To what extent do tools support the different phases of MLOps?
\end{tabular}

\vspace{2mm}
where we aim at understanding which tools can be used in the different phases of the MLOps process.


To answer such RQ in an objective, systematic, and reproducible manner, we adopted the Systematic Multivocal Literature Review (MLR) methodology following the guidelines proposed by Garousi et al.~\cite{Garousi18a}.

The MLR process includes both peer-reviewed as well as gray literature and the different perspectives between practitioners and academic researchers are taken into account in the results.
MLR classifies contributions as \textit{academic literature} in the case of peer-reviewed papers and as \textit{gray literature} in other types of content like blog posts, white papers, podcasts, etc.

\begin{figure}[ht]
    \centering
    \includegraphics[width=.27\textwidth]{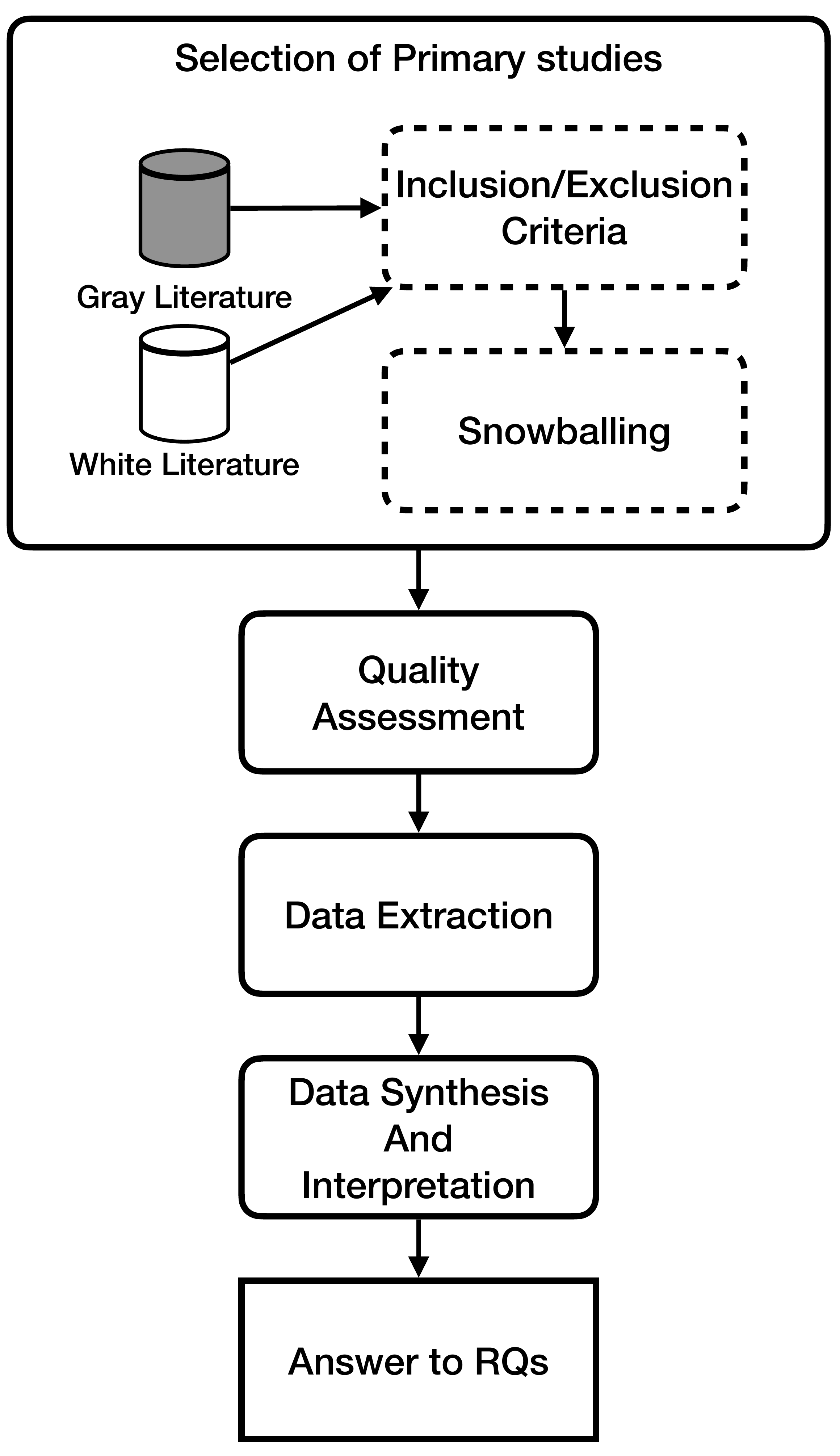}
    \caption{Overview of the followed MLR process}
    \label{fig:mlrprocess}
\end{figure}

The MLR process is composed of four steps, as depicted in Figure \ref{fig:mlrprocess}:
\begin{itemize}
    \item \textit{Selection of primary studies} from the gray and white literature;
    \item \textit{Quality Assessment} of the selected gray literature; 
    \item \textit{Data extraction} of the information needed to answer our RQ;
\begin{itemize}
    \item Data Extraction from the primary studies 
    \item Data Extraction from the tools' websites
\end{itemize}
\item \textit{Tools Selection } 
\item \textit{Data synthesis} of the extracted results 
\end{itemize}


\subsection{Selection of Primary Studies}
\label{sec:Selection}
The first step for selecting the Primary Studies (PS) is the search string identification that will be adopted in the academic bibliographic sources and the gray literature source engines. 

To obtain a high recall and include as many papers as possible, we extended the search string used by Recupito et al. \cite{Recupito2022}:

\begin{center}
\textbf{(``ml-ops'' OR ``ml ops'' OR mlops OR ``machine learning ops'' OR ``machine learning operations'' OR dataops) \\
AND \\
(tool OR application) \\
AND \\
(lifecycle OR pipeline OR platform OR workflow)}
\end{center}

The search terms were applied to all the fields (i.e. title, abstract, and keywords) to include as many works as possible. We adopted the same search terms for retrieving gray literature from online sources and for white literature from academic bibliographic sources.

\textbf{Peer-reviewed literature search}. We considered the papers indexed by six bibliographic sources:
\begin{itemize}
    \item ACM digital Library\footnote{\url{https://dl.acm.org}}
    \item IEEEXplore Digital Library\footnote{\url{https://ieeexplore.ieee.org}}
    \item Science Direct\footnote{\url{https://www.sciencedirect.com}}
    \item Scopus\footnote{\url{https://www.scopus.com}}
    \item Google Scholar\footnote{\url{https://scholar.google.com}}
    \item Springer link\footnote{\url{https://link.springer.com/}}
\end{itemize}
Both searches (gray and white literature) were conducted in November 2022.

\textbf{Gray literature search}. 
We performed the search using four search engines: 

\begin{itemize}
    \item Google Search\footnote{\url{https://www.google.com/}} 
    \item Twitter\footnote{\url{https://twitter.com/}}
    \item Reddit\footnote{\url{https://www.reddit.com/}} 
    \item Medium\footnote{\url{https://medium.com}}
\end{itemize}

The search results consisted of books, blog posts, forums, websites, videos, white-paper, frameworks, and podcasts.

\textbf{Snowballing}. Snowballing refers to using the reference list of a paper or the citations to the paper to identify additional primary studies \cite{Wohlin2014}. We applied backward snowballing to the academic literature to identify relevant primary studies from the references of the selected sources. Moreover, we applied backward-snowballing for the gray literature following outgoing links of each selected source. 

\textbf{Application of inclusion and exclusion criteria}. Based on guidelines for Systematic Literature Reviews \cite{Kitchenham2007}, we defined \textit{inclusion and exclusion} criteria (Table \ref{tab:inclusionExclusion}). 
We considered less restrictive inclusion criteria to enable the inclusion of a more comprehensive tools set.

Before applying the inclusion and exclusion criteria, we tested their applicability~\cite{Kitchenham2013} on a subset of 25 PSs randomly selected from the retrieved ones.
The final set of inclusion and exclusion criteria is summarized in Table~\ref{tab:coding}.

In order to screen each paper, we assigned two researchers to independently review them. To ensure fairness, we mixed up the assignments and made sure each researcher had a similar number of papers to review with other members of the team. In case of disagreement, a third author was brought in to reach a consensus. This was done to improve the reliability of our study. Finally, to evaluate the inter-rater agreement before involving the third author, we calculated Cohen's kappa coefficient. 
These results are  documented in the replication package~\footref{Package}.

\begin{table*}[h]
\centering
\footnotesize
    \caption{Inclusion/Exclusion Criteria for Selection }
    \label{tab:coding}
    \begin{tabular}{l|p{15cm}}
    \hline 
& \textbf{Criteria}\\ \hline
\multirow{1}{*}{\textbf{Inclusion}} & PS mentioning tools that can be used to create or compose MLOps pipelines \\ \hline
\multirow{6}{*}{\textbf{Exclusion}} & - Not in English  \\
& - Duplicated (post summarizing other website)\\
& - Out of topic (using the terms for other purposes)\\
& - Non peer-reviewed papers \\
& - Research Plans, Roadmaps, Vision Papers, Master Thesis \\
& - Less than 750 followers for the author (only for medium - gray literature)*\\ 
\hline 
\textbf{Tools Selection} & \textbf{Criteria}\\ \hline
\multirow{1}{*}{\textbf{Inclusion}} & Commercial and Open Source Tools \\\hline
\multirow{9}{*}{\textbf{Exclusion}} & - Not downloadable  \\
& - No website available/reachable\\ 
& - Not in English\\
& - Duplicated\\
& - Hardware\\
& - Not Reliable (ex. Not enough Stars in GitHub)\\ 
& - Not essentials (ex. Training Orchestration, Explainable AI, FeatureStore)\\ 
& - Addons\\
& - Tools breaking the automatization (Data Generation, Data Labeling)\\
\hline
\multicolumn{2}{l}{* we set up this threshold because some account have been opened only to promote tools or authors did not gain enough followers due} \\
\multicolumn{2}{l}{to their previous publications} \\
\end{tabular}
\label{tab:inclusionExclusion}
\end{table*}

\subsection{Quality Assessment of the Gray Literature}
\label{sec:Quality}
Differently than peer-reviewed literature, gray literature does not go through a formal review process, and therefore its quality is less controlled. To evaluate the credibility and quality of the selected gray literature sources and to decide whether to include a gray literature source or not, we 
followed the recommendation of Garousi et al. guidelines~\cite{Garousi18a}, considering the authority of the producer, the applied methodology, objectivity, date, novelty, and impact. 

The first two authors assessed each source using the aforementioned criteria, with a binary or three-point Likert scale, depending on the criteria itself. In case of disagreement, we discussed the evaluation with the third author that helped to provide the final assessment.




\subsection{Data Extraction}
As our goal is to characterize information from MLOps tools, we need to get these pieces of information directly from the tools' websites. 
Therefore, the data extraction process is composed of two steps: 
\begin{itemize}
    \item[(PE)]\textit{Extraction of the list of tools from the primary studies} (PSs) that satisfied the quality assessment criteria. 
    \item[(TE)] \textit{Extraction of the information from the tools list}. In this case, we extracted the information directly from the official website portals.
\end{itemize}

Based on the RQ, we extracted the information in a review spreadsheet. 
The data extraction form, together with the mapping of the information needed to answer the RQ, is summarized in Table~\ref{tab:ext}.

\begin{table}[]
    \centering
    \footnotesize
    \caption{Data Extraction }
    \label{tab:ext}
    \begin{tabular}{l|p{4.2cm}|l} \hline
 	\textbf{Info}	&	\textbf{Description}	& \textbf{Step	}\\\hline
 Tool Name	&	Name of the tool & \multirow{2}{*}{PE}\\
 Tool Url &  \\\hline
 	\multirow{3}{*}{Phase}	&	Main DevOps phase covered by the tool (Plan, Code, Build, Test, Release, Deploy, Operate, Monitor) & \multirow{8}{*}{TE}\\\cline{1-2}
	\multirow{2}{*}{Purpose}	&	Main purpose of the tool &  \\
 & (e.g. Continuous Training) &  \\\cline{1-2}
 	Alternatives	&	Main competitors of the tool		\\\cline{1-2}

\hline
\end{tabular}
\end{table}



The data extraction process was conducted following the guidelines for qualitative analysis proposed by Wohlin et al.~\cite{WohlinExperimentation}. 
Two researchers extracted all the information. In case of disagreement, we discussed the results  involving a third author. The discussion was conducted until the disagreement was solved.

\subsection{Tool Selection}

From the list of tools extracted in the previous step, we need to identify the final set to answer our RQ. Therefore, we applied a similar process of the one adopted in the papers selection phase (Section~\ref{sec:Selection}), filtering the tools based on a set of inclusion and exclusion criteria. 

Before applying the inclusion and exclusion criteria, we tested their applicability~\cite{Kitchenham2013} on a subset of 10\% of tools randomly selected from the retrieved ones.
The final set of inclusion and exclusion criteria is reported in Table \ref{tab:inclusionExclusion}.

As well as for the PS selection, the selection of the tools was performed by two researchers, that independently reviewed them. Also in this case, in case of disagreement, a third author was involved to reach a consensus. The Cohen's kappa coefficient has been calculated also in this case, to confirm the inter-rater agreement. 
These results are  documented in the replication package~\footref{Package}.


\subsection{Conducting the review}

From the Search process, we retrieved a total of 783 unique PS (after the exclusion of 137 duplicated): 497 PS from the gray literature and 287 PS from the white literature. The snowballing process enabled to include 30 more sources (\SnowballingWhite from white literature and \SnowballingGray from gray literature). 

After the application of inclusion and exclusion criteria, we selected \SelectedPS PS (51 from the white literature and 203 from the gray literature).
The application of the inclusion criteria performed by the author had an almost perfect agreement (Cohen's kappa = 0.870). 

The application of the quality assessment process to the gray literature PS, resulted in the exclusion of 1 PS. 

As a result, we included a total of \SelectedPS PS, of which \SelectedGray PS from the gray literature and \SelectedWhite PS from the white literature. 

From the data extraction process on the \SelectedPS selected PS we obtained \SelectedTools MLOps Tools. 

The application of the inclusion and exclusion criteria resulted in an almost perfect agreement (The Cohen's Kappa coefficient = 0.812). As a result, we obtained a final set of \SelectedTools as reported in Table \ref{tab:selectedTools}.

\subsection{Verifiability and Replicability}
\noindent To allow our study to be replicated, we have published the complete raw data in the replication package.\footnote{\url{https://figshare.com/s/0b62981a4ed90a93099f}\label{Package}}.


\section{Analysis of the Results}
\label{sec:Results}

\begin{figure*}
\centering
{\includegraphics[width=0.9\textwidth]{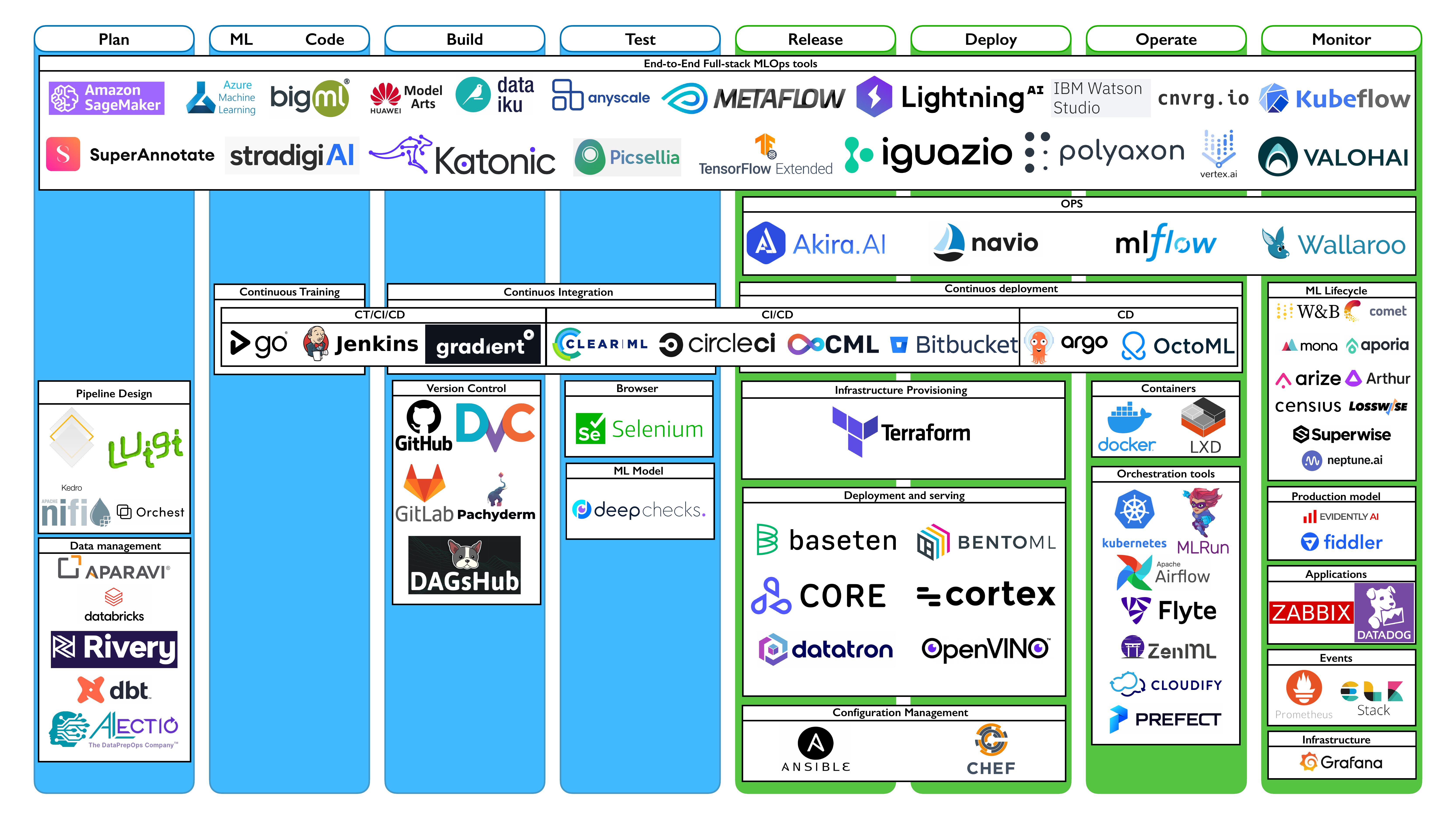}}

\caption{Identified tools and their subcategories to produce MLOps Pipelines}
\label{fig:Pipelines}

\end{figure*}


We extracted characteristics of the identified tools and their alternatives and grouped them into the following subcategories based on the MLOps phases covered by each tool. We identified 5 global categories:
\begin{enumerate}
    \item \textbf{Continuous Development (CD)}: based on the DevOps classic definition, CD includes those tools which create a single software package or a service. In order to extend this to MLOps we also include those tools which create a system that automatically deploys another service \cite{GoogleMLOps}.
    \item \textbf{Continuous Integration (CI)}: based on the DevOps classic definition, CI includes the tools used for testing and validating code and components. In order to extend this to MLOps we also include tools for testing and validating data and models \cite{GoogleMLOps}.
    \item \textbf{Continuous Testing (CT)}: tools used to automatically retrain and serve the models \cite{GoogleMLOps}.
    \item \textbf{End-to-End Full Stack MLOps tools}: includes tools that can be used to create a full pipeline singularly.
    \item \textbf{Operations (OPS)}: based on the DevOps classic definition, OPS includes those tools which manage the phases of Release, Deploy, Operate, and Monitor.
\end{enumerate}

\vspace{2mm}
Then, we refined our classification considering the phases described in Section \ref{sec:Background}. 

\vspace{2mm}
\textbf{Plan}. We identified 2 subcategories: 
\begin{enumerate}
\item \textit{Data Management}: tools responsible to manage the data used to train the ML models.
\item \textit{Pipeline Design}: tools used to create pipelines. These can include dependency resolution, workflow management, and visualization.
\end{enumerate}


\textbf{Build}. As subcategory, we identified \textit{Version Control}: tools responsible for tracking and managing changes to code

\textbf{Test}. We identified 2 subcategories: 
\begin{enumerate}
\item \textit{Code}: code-based test tools. 
\item \textit{ML Model}: tools used for testing the ML models and the data used to train it.
\end{enumerate}

\textbf{Release/Deploy}. We identified 3 subcategories: 
\begin{enumerate}
\item \textit{Configuration Management}: tools used to perform configuration management i.e, establishing consistency of a product’s attributes throughout its life.
\item \textit{Deployment and Serving}: tools used to automate the process of deploying machine learning models into production. The process automates many sub-tasks including model serving, a technique for integrating an ML model into a software system.
\item \textit{Infrastructure Provisioning}: tools used for the process of provisioning or creating infrastructure resources.

\end{enumerate}

\textbf{Operate}. We identified 2 subcategories: 
\begin{enumerate}
\item \textit{Containers}: tools used to create containers. A container is defined as a ``standard unit of software that packages up code and all its dependencies so the application runs quickly and reliably from one computing environment to another'' \cite{Docker}.
\item \textit{Orchestration Tools}: tools used to automate the process of coordination of containers. This includes start and stop, schedule and execution of tasks, and recovery processes.
\end{enumerate}

\textbf{Monitor}. We identified 5 subcategories: 
\begin{enumerate}
\item \textit{Applications}: tools used to examine and investigate performance data of created applications.
\item \textit{Events}: tools used for event monitoring and providing alerting functionalities. 
\item \textit{Infrastructure}: tools used to monitor infrastructure nodes.
\item \textit{ML Lifecycle}: tools used for experiment tracking, dataset versioning, and model management of ML models.
\item \textit{Production Model}: tools used to monitor the deployed ML model in order to avoid models breaking or degrading in production.

\end{enumerate}

\begin{figure}[]
\centering
\begin{minipage}{.5\textwidth}
  \centering
  \includegraphics[width=0.7\linewidth]{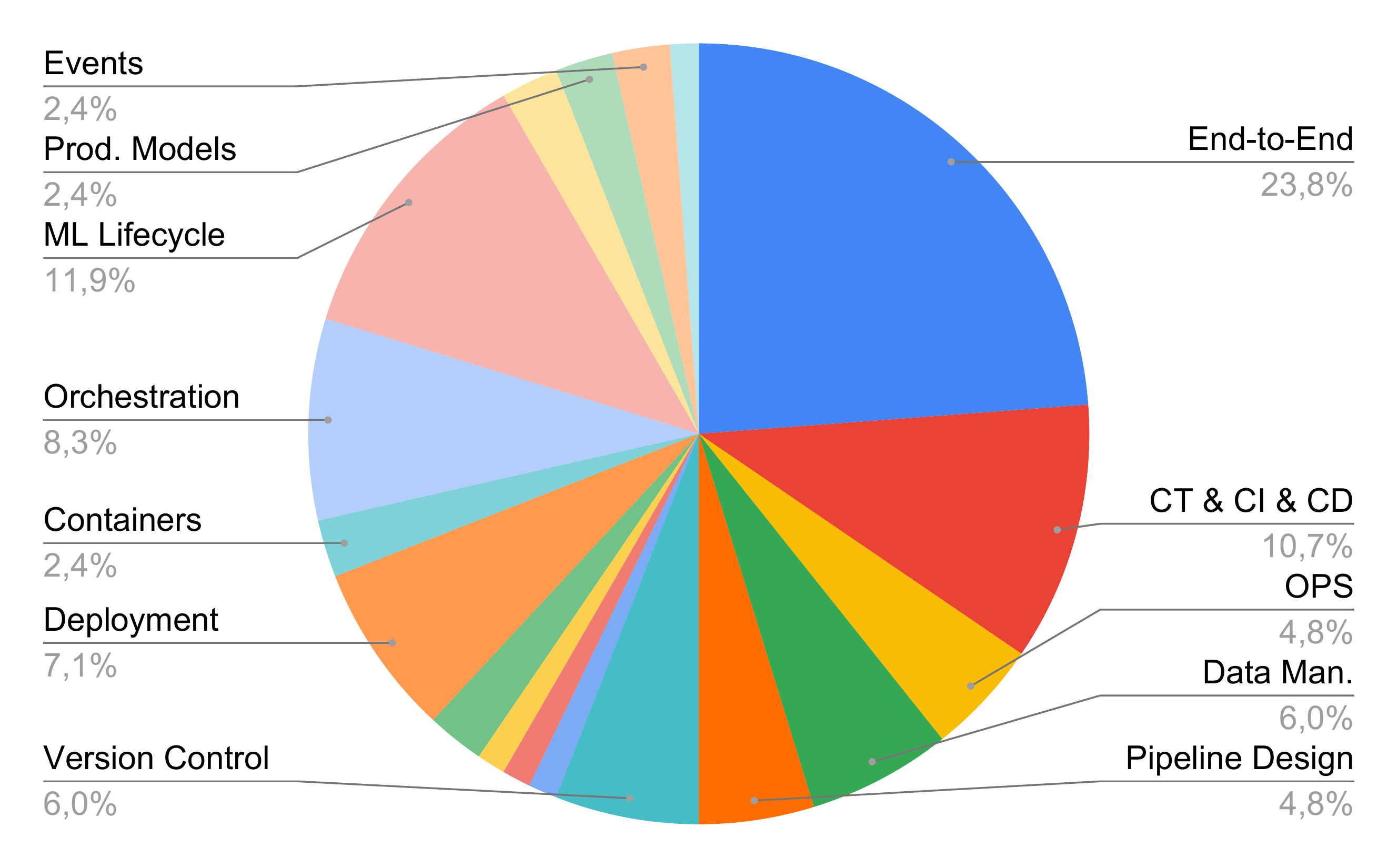}
  \caption{Selected Tools by Subcategory}
  \label{fig:toolbysub}
\end{minipage}%
 \\
\begin{minipage}{.5\textwidth}
  \centering
  \includegraphics[width=0.7\linewidth]{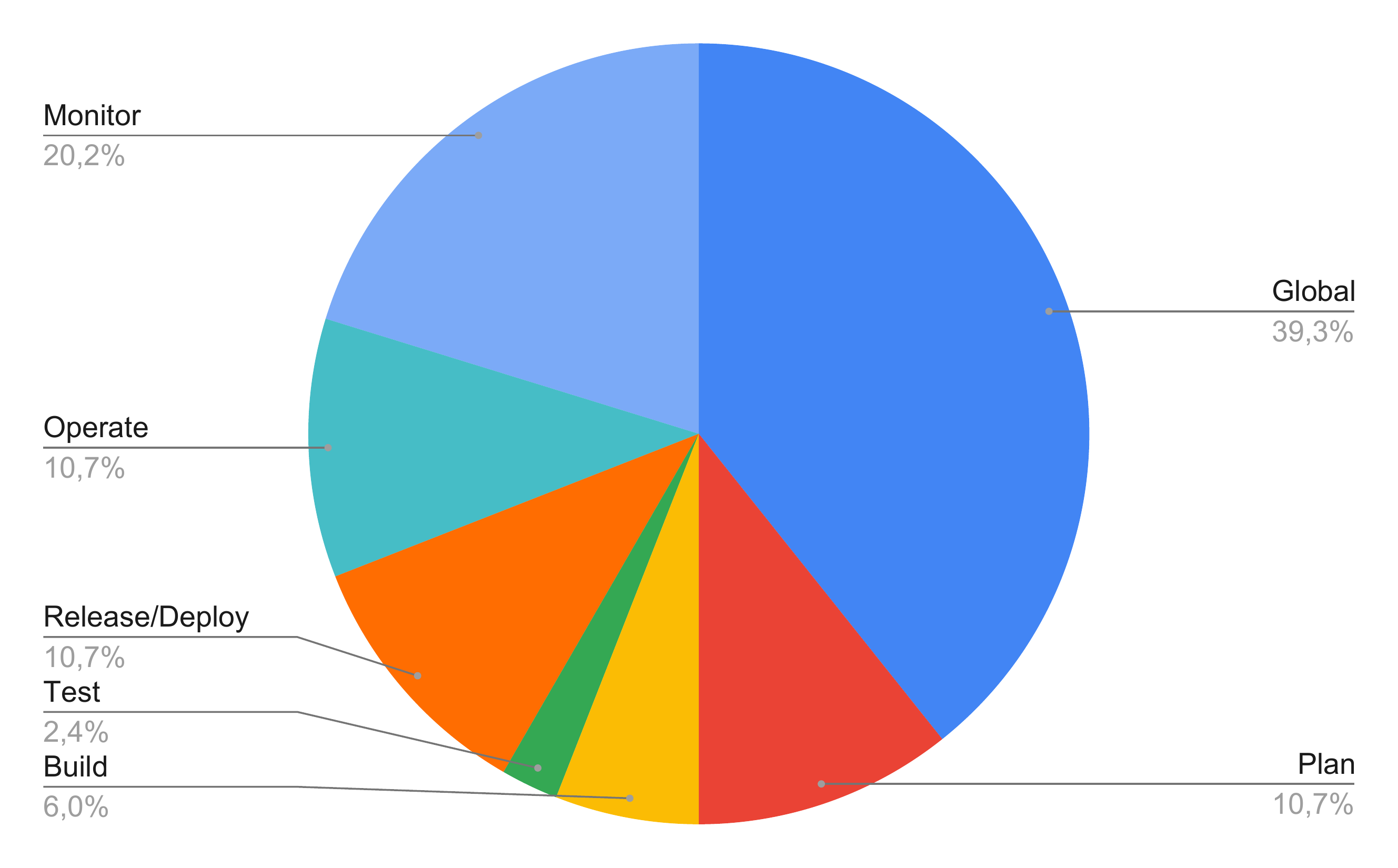}
  \caption{Selected Tools by Phase}
  \label{fig:toolbyphase}
\end{minipage}
\label{fig:papertypeandyear}
\end{figure}

The final result of the categorization of the different MLOps tools is illustrated in Figure~\ref{fig:Pipelines}. This resulted in \SelectedTools different tools being divided into 20 different categories. The amount of different tools for each subcategory is reported in Figure~\ref{fig:toolbysub} and Table~\ref{tab:categories} where the tools for CT, CI, and CD have been grouped. 

Figure~\ref{fig:toolbysub} shows the distribution of tools based on the phase of the MLOps pipeline they support. Also, in this case, CT, CI, and CD have been grouped.

\begin{table}[h]
\centering
\footnotesize
    \caption{Tools and Categories}
    \label{tab:categories}
    \begin{tabular}{p{1.5cm}|l|r}
    \hline 
\textbf{Category} & \textbf{Subcategory}& \textbf{ \# tools}\\ 
        \hline
\multirow{3}{*}{\textbf{Global}} & - CT \& CI \& CD & 9\\
         & - End-to-End Full Stack MLOps tools & 20 \\
         & - OPS & 4 \\
        \hline
\multirow{2}{*}{\textbf{Plan}} & - Data Management & 5 \\
& - Pipeline Design & 4\\
        \hline
\textbf{Build}& Version Control & 5\\
        \hline
\multirow{2}{*}{\textbf{Test}} &- Browser & 1 \\
& - ML Model & 1 \\
\hline
\textbf{Release \&} & - Deployment and Serving & 6\\
\textbf{Deploy} & - Infrastructure Provisioning & 1\\
\hline
 & - Configuration Management & 2\\  
\multirow{3}{*}{\textbf{Operate}} & - Containers & 2 \\
& - Orchestration tools & 7 \\
\hline
\multirow{5}{*}{\textbf{Monitor}} & - Applications & 2 \\
& - Events & 2\\
& - Infrastructures & 1\\
& - ML Lifecycle & 10\\
& - Production Models & 2\\
        \hline 
    \end{tabular}
    \label{tab:incexc}
\end{table}

\newpage
\section{Discussion and Implications}
\label{sec:Discussion}

The results of the study provide a number of implications and actionable points that are worth discussing further. 

\smallskip
\begin{description}[leftmargin=0.5cm]
\item \textbf{On the existence of MLOps tools.} 

First and foremost, our findings revealed the existence of \SelectedTools MLOps tools. Such a large amount of instruments were mostly made available by tool vendors and practitioners---as indicated by the number of tools identified within the grey literature review analysis. 
On the one hand, this highlights the ever-increasing attention that practitioners have with respect to MLOps activities, further motivating the research around the matter and the need for a comprehensive analysis of how to combine those tools, as preliminarily done within the scope of our work. 
On the other hand, we point out a noticeable lack of instruments proposed or empirically investigated by our research community.
In particular, both the comprehensive list of MLOps currently available and the mapping onto the MLOps pipeline, which are the key outcomes of our work, might serve as a basis for further investigations into various angles of the problem, like the actionability of these tools in practice, the information needs required by practitioners to effectively use those tools, or the most effective strategies to combine multiple tools. 

\smallskip
{\emph{Implication \#1.}}
\textit{Our study may therefore represent the first step toward increased awareness of the impact that the empirical software engineering research field may have on the development of new MLOps solutions. }

\smallskip 
\item \textbf{On the compatibility of MLOps tools.}

  The  \textit{``End-to-End  MLOps tools''} usually provide a predefined tool pipeline. However, in some cases, it is possible to replace their tools with alternative ones. Such tools need to cover all of the phases of the MLOps pipeline. For an optimal pipeline generation, not involving an  \textit{``End-to-End  MLOps tools''} it is therefore needed the use a tool for each one of the phases in the \textit{``Dev''} phase and at least one tool for the \textit{``Ops''} phases \cite{OPS}. The reason behind it is that while including multiple tools for the \textit{``Ops''} phases might produce a valuable and improved output, the inclusion of multiple tools from the same phase of the \textit{``Dev''} creates redundancy and might break the pipeline automation.
According to our mapping exercise, the tools that emerged from the systematic study seem to be compatible, meaning that they offer orthogonal pieces of information throughout the MLOps pipeline. 

\smallskip
{\emph{Implication \#2.}}
\textit{We believe that this result might enable several future investigations on the matter: questions concerned with the information flows required to sequentially use tools or the redundant/contrasting information that multiple tools might provide to practitioners are indeed still open and neglected by our research community.}

\smallskip
\item \textbf{On the creation of a taxonomy.}

Last but not least, the mapping exercise conducted in our work may have implications for tool vendors and practitioners. The former might find a comprehensive overview of the state of the art/practice which might potentially serve as a basis for developing novel instruments covering the steps of the MLOps pipeline that currently suffer from a lower representation. In addition, the overview might also enable further considerations, like an informed analysis of the features provided by those tools, which in turn can lead tool vendors to improve the existing instruments and/or design additional support systems. 

\smallskip
{\emph{Implication \#3.}}
\textit{Practitioners can use our taxonomy to tune their own MLOps pipelines by uncovering alternative tools they were unaware of or experimenting with combinations of tools that might help them cover the entire MLOps pipeline. }

\end{description}
\section{Threats to Validity}
\label{sec:Threats}
The results of our study may be affected by various sources of bias or error, including inaccuracies in data extraction, limitations in the scope of the literature review, subjectivity in the definition and application of inclusion and exclusion criteria. In this section, we address these potential threats by outlining the strategies we employed to mitigate them, as per the guidelines outlined in ~\cite{WohlinExperimentation}.

\textbf{Construct validity}. Construct validity concerns the extent to which the study's object of investigation accurately reflects the theory behind the study, according to a reference. 

The research questions and classification schema used in the study may be subject to this type of threat. To minimize this risk, the authors independently reviewed and then discussed the research questions. As for the classification schema adopted, we selected a standard classification of DevOps. 

\textbf{Internal Validity}. The source selection approach adopted in this work is described in Section~\ref{sec:Selection}. In order enable the replicability of our work, we carefully identified and reported bibliographic sources adopted to identify the peer-review literature, search engines, adopted for the gray literature,  search strings as well as inclusion and exclusion criteria.
Possible issues in the selection process are related to the selection of search terms that could have lead to a non complete set of results. 
To mitigate this risk, we broadened the search string adopted by Recupito et al. including possible synonyms for the term MLOps. 
To overcome the limitation of the search engines, we queried the academic literature from six bibliographic sources, while we included the gray literature from Google, Medium Search, Twitter Search and Reddit Search. Additionally, we applied a snowballing process to include all the possible sources.
The application of inclusion and exclusion can be affected by researchers’ opinion  and experience. To mitigate this threat, all the sources were evaluated by at least two authors independently. Moreover, to evaluate the quality of the inter-rater agreement we calculated the Cohen's Kappa coefficient, obtaining an almost perfect agreement. We believe that such high agreement is due to the easiness of the application of the inclusion and exclusion criteria, and in particular it was easy to exclude papers not containing any reference to any tool.

\textbf{Conclusion validity}.
Conclusion validity is related to the reliability of the conclusions drawn from the results~\cite{WohlinExperimentation}.
To ensure the reliability of our treatments, we did not define the terminology ourselves, but we based our classification on the existing DevOps and MLOps steps. 
Moreover, all primary sources were reviewed by at least two authors to mitigate bias in data extraction and each disagreement was resolved by consensus, involving a third author.

\textbf{External Validity.}
External validity is related to the generalizability of the  results of our multivocal literature review.  In this study, we map the literature on MLOps Tools, considering both the academic and the gray literature. However, we cannot claim to have screened all the possible literature, since some documents might have not been properly indexed, or possibly copyrighted or, even not freely available.
 
\section{Related Work}
\label{sec:RW}
Several studies highlight the current state of tools that assist the ML pipeline. On the basis of feature analysis, Idowu et al. \cite{IdowuAssetManagement}  give a comparison of 30 ML asset management solutions. The findings demonstrate that ML-enabled systems with tracking, exploring, and retrieving activities for experiment reproducibility are supported by the state of practice and the state of research. By evaluating the 26 MLOps tools that may be supported by the ML Pipeline and examining their potential support, Ruf et al. \cite{RufDemystifying} show the potential of MLOps. The findings show that none of the MLOps solutions are fully automated enough to support the MLOps workflow, and that several products have capabilities that overlap and offer the same support.
Kolltveit et al. \cite{KolltveitOperationalizing} conducted an SLR to investigate the actual tools that assist in model operationalization, with the purpose of identifying feature gaps. The review's findings showed that managing the edge deployment of ML models and implementing features like dynamic model switching and continuous model monitoring present real challenges for MLOps. Testi et al.\cite{TestiTaxonomy}  conducted a review to establish an MLOps taxonomy that outlines the approaches and processes used to establish an ML pipeline. The authors give a summary of the advantages and disadvantages of employing a selected tool in a specific stage of the ML Pipeline based on the collection of tools they have uncovered.
Finally, Recupito et al. \cite{Recupito2022} presented a multivocal
literature review of the several MLOps tools and features
present in the state of the art and the state of the practice.
After extracting a set of 13 MLOps tools, a feature analysis
and comparison of aspects that involve the whole machine
learning pipeline, aspects that involve the model management
and aspects that involve data management.

The most of reviews examined in the state-of-the-art emphasize the need to bridge the gap between research and practice. The results of this research take into account 84 tools that practitioners employ to automate and maintain ML systems.
Additionally, we expanded the research to the current state of MLOps tools by examining their compatibilities, giving practitioners the opportunity to select the ideal set of solutions for employing MLOps in ML systems.

\section{Conclusion}
\label{sec:Conclusion}

In this work, we performed a Multivocal Literature Review to classify the MLOps tools in the DevOps process, and to identify possible incompatibilities among tools. 

We extracted \SelectedTools tools from \SelectedPS primary studies (\SelectedGray PS from the gray literature and \SelectedWhite PS from the white literature).
We finally proposed a graphical representation for the MLOps tools into the DevOps process, that will be useful to researchers and practitioners to have a quick overview on the existing MLOps tools. 
It is interesting to note that the vast majority of end-to-end MLOps platforms do not commonly enable to integrate other tools, while no incompatibilities were found among other tools. 

As a result, it is possible to combine all the tools from different DevOps stages into a seamless pipeline. 

We are planning to extend this work conducting an industrial survey to investigate the most common pipelines adopted  
in industry and the perceived benefits, issues, and usefulness of each tool.

\section*{Acknowledgments}
This work is partially funded by  the IndustryX project (Business Finland), the MuFAno project from the Academy of Finland (grant n. 349488), the EMELIOT project (Italy, MUR -  PRIN 2020 program, contract 2020W3A5FY) and the  
Swiss National Science Foundation through SNF Projects No. PZ00P2\_186090.

\section*{CRediT authorship contribution statement}
\textbf{Sergio Moreschini:} Conceptualization, Investigation, Data Curation, Writing - Original Draft, Writing - Review \& Editing, Visualization.

\textbf{Gilberto Recupito:} Investigation, Data Curation, Writing - Review \& Editing.

\textbf{Valentina Lenarduzzi:} Writing - Original Draft, Writing - Review \& Editing, Visualization.

\textbf{Fabio Palomba:} Supervision Reviewing, Editing. 

\textbf{David H\"astbacka:} Conceptualization, Methodology, Supervision Reviewing, Editing, Funding Acquisition.

\textbf{Davide Taibi:} Conceptualization, Methodology, Supervision Reviewing, Editing, Funding Acquisition.

\bibliographystyle{model1-num-names}
\bibliography{sample}

\begin{thebibliography}{16}
\expandafter\ifx\csname natexlab\endcsname\relax\def\natexlab#1{#1}\fi
\providecommand{\bibinfo}[2]{#2}
\ifx\xfnm\relax \def\xfnm[#1]{\unskip,\space#1}\fi
\bibitem[{Recupito et~al.(2022)Recupito, Pecorelli, Catolino, Moreschini,
  Di~Nucci, Palomba, and Tamburri}]{Recupito2022}
\bibinfo{author}{G.~Recupito}, \bibinfo{author}{F.~Pecorelli},
  \bibinfo{author}{G.~Catolino}, \bibinfo{author}{S.~Moreschini},
  \bibinfo{author}{D.~Di~Nucci}, \bibinfo{author}{F.~Palomba},
  \bibinfo{author}{D.~A. Tamburri},
\newblock \bibinfo{title}{A multivocal literature review of mlops tools and
  features},
\newblock \bibinfo{journal}{Euromicro Conference on Software Engineering and
  Advanced Applications}  (\bibinfo{year}{2022}).
\bibitem[{Carver et~al.(2014)Carver, Juristo, Baldassarre, and
  Vegas}]{CarverReplication2014}
\bibinfo{author}{J.~Carver}, \bibinfo{author}{N.~Juristo},
  \bibinfo{author}{M.~T. Baldassarre}, \bibinfo{author}{S.~Vegas},
\newblock \bibinfo{title}{Replications of software engineering experiments},
\newblock \bibinfo{journal}{Empirical Software Engineering}
  (\bibinfo{year}{2014}) \bibinfo{pages}{267–276}.
\bibitem[{Bass et~al.(2015)Bass, Weber, and Zhu}]{bass2015devops}
\bibinfo{author}{L.~Bass}, \bibinfo{author}{I.~Weber},
  \bibinfo{author}{L.~Zhu}, \bibinfo{title}{DevOps: A software architect's
  perspective}, \bibinfo{publisher}{Addison-Wesley Professional},
  \bibinfo{year}{2015}.
\bibitem[{Moreschini et~al.(2022)Moreschini, Lomio, Hästbacka, and
  Taibi}]{Moreschini22}
\bibinfo{author}{S.~Moreschini}, \bibinfo{author}{F.~Lomio},
  \bibinfo{author}{D.~Hästbacka}, \bibinfo{author}{D.~Taibi},
\newblock \bibinfo{title}{Mlops for evolvable ai intensive software systems},
\newblock in: \bibinfo{booktitle}{2022 IEEE International Conference on
  Software Analysis, Evolution and Reengineering (SANER)}, pp.
  \bibinfo{pages}{1293--1294}.
\bibitem[{Garousi et~al.(2019)Garousi, Felderer, and Mäntylä}]{Garousi18a}
\bibinfo{author}{V.~Garousi}, \bibinfo{author}{M.~Felderer},
  \bibinfo{author}{M.~V. Mäntylä},
\newblock \bibinfo{title}{{Guidelines for including grey literature and
  conducting multivocal literature reviews in software engineering}},
\newblock \bibinfo{journal}{Information and Software Technology}
  \bibinfo{volume}{106} (\bibinfo{year}{2019}) \bibinfo{pages}{101--121}.
\bibitem[{Wohlin(2014)}]{Wohlin2014}
\bibinfo{author}{C.~Wohlin},
\newblock \bibinfo{title}{{Guidelines for Snowballing in Systematic Literature
  Studies and a Replication in Software Engineering}},
\newblock in: \bibinfo{booktitle}{International Conference on Evaluation and
  Assessment in Software Engineering}, Ease '14, pp. \bibinfo{pages}{1--10}.
\bibitem[{Kitchenham and Charters(2007)}]{Kitchenham2007}
\bibinfo{author}{B.~Kitchenham}, \bibinfo{author}{S.~Charters},
  \bibinfo{title}{{Guidelines for performing Systematic Literature Reviews in
  Software Engineering}}, \bibinfo{type}{Technical Report}
  \bibinfo{number}{Ebse 2007-001}, Keele University, \bibinfo{year}{2007}.
\bibitem[{Kitchenham and Brereton(2013)}]{Kitchenham2013}
\bibinfo{author}{B.~Kitchenham}, \bibinfo{author}{P.~Brereton},
\newblock \bibinfo{title}{A systematic review of systematic review process
  research in software engineering},
\newblock \bibinfo{journal}{Information {\&} Software Technology}
  \bibinfo{volume}{55} (\bibinfo{year}{2013}) \bibinfo{pages}{2049--2075}.
\bibitem[{Wohlin et~al.(2012)Wohlin, Runeson, Höst, Ohlsson, and
  Regnell}]{WohlinExperimentation}
\bibinfo{author}{C.~Wohlin}, \bibinfo{author}{P.~Runeson},
  \bibinfo{author}{M.~Höst}, \bibinfo{author}{M.~C. Ohlsson},
  \bibinfo{author}{B.~Regnell}, \bibinfo{title}{{Experimentation in Software
  Engineering.}}, \bibinfo{year}{2012}.
\bibitem[{Center(2022)}]{GoogleMLOps}
\bibinfo{author}{G.~A. Center}, \bibinfo{title}{Mlops: Continuous delivery and
  automation pipelines in machine learning},
  \bibinfo{howpublished}{\url{https://cloud.google.com/architecture/mlops-continuous-delivery-and-automation-pipelines-in-machine-learning\#devops\_versus\_mlops}},
  \bibinfo{year}{2022}.
\bibitem[{Docker(2022)}]{Docker}
\bibinfo{author}{Docker}, \bibinfo{title}{What is a container?},
  \bibinfo{howpublished}{\url{https://www.docker.com/resources/what-container/}},
  \bibinfo{year}{2022}.
\bibitem[{Felix(2022)}]{OPS}
\bibinfo{author}{A.~Felix}, \bibinfo{title}{How to monitor jenkins metrics
  using prometheus \& grafana?},
  \bibinfo{howpublished}{\url{https://medium.com/@AnnFelix/how-to-monitor-jenkins-metrics-using-prometheus-grafana-152a98d6c7a6}},
  \bibinfo{year}{2022}.
\bibitem[{Idowu et~al.(2022)Idowu, Str\"{u}ber, and
  Berger}]{IdowuAssetManagement}
\bibinfo{author}{S.~Idowu}, \bibinfo{author}{D.~Str\"{u}ber},
  \bibinfo{author}{T.~Berger},
\newblock \bibinfo{title}{Asset management in machine learning:
  State-of-research and state-of-practice},
\newblock \bibinfo{journal}{ACM Comput. Surv.} \bibinfo{volume}{55}
  (\bibinfo{year}{2022}).
\bibitem[{Ruf et~al.(2021)Ruf, Madan, Reich, and
  Ould-Abdeslam}]{RufDemystifying}
\bibinfo{author}{P.~Ruf}, \bibinfo{author}{M.~Madan},
  \bibinfo{author}{C.~Reich}, \bibinfo{author}{D.~Ould-Abdeslam},
\newblock \bibinfo{title}{Demystifying mlops and presenting a recipe for the
  selection of open-source tools},
\newblock \bibinfo{journal}{Applied Sciences} \bibinfo{volume}{11}
  (\bibinfo{year}{2021}).
\bibitem[{Kolltveit and Li(2022)}]{KolltveitOperationalizing}
\bibinfo{author}{A.~B. Kolltveit}, \bibinfo{author}{J.~Li},
\newblock \bibinfo{title}{Operationalizing machine learning models - a
  systematic literature review},
\newblock in: \bibinfo{booktitle}{2022 IEEE/ACM 1st International Workshop on
  Software Engineering for Responsible Artificial Intelligence (SE4RAI)}, pp.
  \bibinfo{pages}{1--8}.
\bibitem[{Testi et~al.(2022)Testi, Ballabio, Frontoni, Iannello, Moccia, Soda,
  and Vessio}]{TestiTaxonomy}
\bibinfo{author}{M.~Testi}, \bibinfo{author}{M.~Ballabio},
  \bibinfo{author}{E.~Frontoni}, \bibinfo{author}{G.~Iannello},
  \bibinfo{author}{S.~Moccia}, \bibinfo{author}{P.~Soda},
  \bibinfo{author}{G.~Vessio},
\newblock \bibinfo{title}{Mlops: A taxonomy and a methodology},
\newblock \bibinfo{journal}{IEEE Access} \bibinfo{volume}{10}
  (\bibinfo{year}{2022}) \bibinfo{pages}{63606--63618}.

\end{thebibliography}
\begin{table*}
\centering
\scriptsize
\caption{The Selected MLOps Tools}
\label{tab:selectedTools}
   \resizebox{\textwidth}{!}{
\begin{tabular}{p{4cm}lllp{1.1cm}p{1.1cm}p{1.1cm}} \\
\hline

\textbf{Tool Name}          & \textbf{Url}                                                                  & \textbf{Category  }            & \textbf{Subcategory}      & \textbf{\#White} & \textbf{\#Gray} & \textbf{\#Total} \\
\hline
DAGsHub	&	https://dagshub.com/	&	Build	&	Version Control	&	1	&	10	&	11	\\
DVC	&	https://dvc.org/	&	Build	&	Version Control	&	11	&	36	&	47	\\
Github	&	https://github.com	&	Build	&	Version Control	&	6	&	4	&	10	\\
Gitlab	&	https://about.gitlab.com	&	Build	&	Version Control	&	2	&	6	&	8	\\
Pachyderm	&	https://www.pachyderm.com	&	Build	&	Version Control	&	7	&	27	&	34	\\
Argo	&	https://argoproj.github.io/	&	Global	&	CD	&	3	&	7	&	10	\\
OctoML	&	https://octoml.ai	&	Global	&	CD	&	1	&	2	&	3	\\
bitbucket	&	https://bitbucket.org/product/	&	Global	&	CI/CD	&	0	&	1	&	1	\\
CircleCI	&	https://circleci.com	&	Global	&	CI/CD	&	1	&	3	&	4	\\
ClearML	&	https://clear.ml	&	Global	&	CI/CD	&	0	&	10	&	10	\\
CML	&	https://cml.dev	&	Global	&	CI/CD	&	2	&	3	&	5	\\
GoCD	&	https://www.gocd.org	&	Global	&	CI/CD/CT	&	1	&	1	&	2	\\
Gradient	&	https://www.paperspace.com/gradientt	&	Global	&	CI/CD/CT	&	0	&	4	&	4	\\
Jenkins	&	https://www.jenkins.io	&	Global	&	CI/CD/CT	&	5	&	6	&	11	\\
Amazon SageMaker	&	https://aws.amazon.com/sagemaker/	&	Global	&	E2E	&	18	&	28	&	46	\\
Anyscale	&	https://www.anyscale.com/platform	&	Global	&	E2E	&	1	&	1	&	2	\\
Azure ML	&	https://azure.microsoft.com/en-us/products/machine-learning	&	Global	&	E2E	&	13	&	15	&	28	\\
BigML	&	bigml.com	&	Global	&	E2E	&	1	&	0	&	1	\\
CNVRG	&	https://cnvrg.io/	&	Global	&	E2E	&	0	&	3	&	3	\\
Dataiku	&	https://www.dataiku.com/	&	Global	&	E2E	&	1	&	10	&	11	\\
Huawei Cloud ModelArts	&	https://www.huaweicloud.com/intl/en-us/product/modelarts.html	&	Global	&	E2E	&	0	&	1	&	1	\\
Iguazio	&	https://www.iguazio.com/	&	Global	&	E2E	&	4	&	6	&	10	\\
Katonic	&	https://katonic.ai/	&	Global	&	E2E	&	0	&	4	&	4	\\
Kubeflow	&	https://www.kubeflow.org	&	Global	&	E2E	&	13	&	41	&	54	\\
Lightning Ai	&	https://lightning.ai	&	Global	&	E2E	&	0	&	1	&	1	\\
Metaflow	&	https://metaflow.org	&	Global	&	E2E	&	2	&	17	&	19	\\
Picsell.ia	&	https://picsellia.com/	&	Global	&	E2E	&	1	&	1	&	2	\\
Polyaxon	&	https://polyaxon.com/	&	Global	&	E2E	&	2	&	10	&	12	\\
Stradigi AI	&	https://www.stradigi.ai/	&	Global	&	E2E	&	1	&	0	&	1	\\
superannotate MLStudio	&	https://www.superannotate.com/sdk-integration	&	Global	&	E2E	&	2	&	1	&	3	\\
TFX	&	https://www.tensorflow.org/tfx	&	Global	&	E2E	&	11	&	11	&	22	\\
Valohai	&	https://valohai.com/	&	Global	&	E2E	&	4	&	9	&	13	\\
Vertex.ai	&	https://cloud.google.com/vertex-ai	&	Global	&	E2E	&	8	&	16	&	24	\\
Watson Studio	&	https://www.ibm.com/cloud/watson-studio	&	Global	&	E2E	&	2	&	7	&	9	\\
Akira ai	&	https://www.akira.ai/mlops-platform/	&	Global	&	OPS	&	0	&	2	&	2	\\
MLflow	&	https://mlflow.org/	&	Global	&	OPS	&	15	&	55	&	70	\\
navio	&	https://www.craftworks.ai/navio	&	Global	&	OPS	&	0	&	1	&	1	\\
Wallaro	&	https://www.wallaroo.ai	&	Global	&	OPS	&	0	&	1	&	1	\\
Datadog	&	https://www.datadoghq.com/	&	Monitor	&	Applications	&	1	&	3	&	4	\\
Zabbix	&	https://www.zabbix.com	&	Monitor	&	Applications	&	1	&	1	&	2	\\
ELK	&	https://www.elastic.co/what-is/elk-stack	&	Monitor	&	Events	&	3	&	3	&	6	\\
Prometheus	&	https://prometheus.io/	&	Monitor	&	Events	&	3	&	1	&	4	\\
Grafana	&	https://grafana.com	&	Monitor	&	Infrastructure	&	3	&	3	&	6	\\
Aporia	&	https://www.aporia.com	&	Monitor	&	ML Lifecycle	&	1	&	2	&	3	\\
Arize AI	&	https://arize.com	&	Monitor	&	ML Lifecycle	&	4	&	7	&	11	\\
Arthur AI	&	https://www.arthur.ai/	&	Monitor	&	ML Lifecycle	&	1	&	2	&	3	\\
Censius AI	&	https://censius.ai	&	Monitor	&	ML Lifecycle	&	0	&	2	&	2	\\
Comet	&	https://www.comet.ml/	&	Monitor	&	ML Lifecycle	&	2	&	17	&	19	\\
LossWise	&	https://losswise.com/	&	Monitor	&	ML Lifecycle	&	3	&	0	&	3	\\
Mona Labs	&	https://www.monalabs.io/	&	Monitor	&	ML Lifecycle	&	1	&	0	&	1	\\
Neptune.ai	&	https://neptune.ai/	&	Monitor	&	ML lifecycle	&	6	&	18	&	24	\\
Superwise.ai	&	https://superwise.ai	&	Monitor	&	ML Lifecycle	&	2	&	2	&	4	\\
W\&B	&	https://wandb.ai/site	&	Monitor	&	ML Lifecycle	&	3	&	16	&	19	\\
EvidentlyAI	&	https://evidentlyai.com/	&	Monitor	&	Production Model	&	3	&	9	&	12	\\
Fiddler	&	https://www.fiddler.ai/	&	Monitor	&	Production Model	&	3	&	14	&	17	\\
LXD	&	https://linuxcontainers.org/lxd/introduction/	&	Operate	&	Containers	&	9	&	9	&	18	\\
Docker	&	https://www.docker.com	&	Operate	&	Containers	&	1	&	1	&	2	\\
Apache Airflow	&	https://airflow.apache.org/	&	Operate	&	Orchestration	&	7	&	24	&	31	\\
Kubernetes	&	https://kubernetes.io	&	Operate	&	Orchestration	&	1	&	0	&	1	\\
MLrun	&	https://www.mlrun.org	&	Operate	&	Orchestration	&	2	&	12	&	14	\\
Prefect	&	https://www.prefect.io/	&	Operate	&	Orchestration	&	11	&	14	&	25	\\
ZenML	&	https://zenml.io/home	&	Operate	&	Orchestration	&	1	&	8	&	9	\\
Cloudify	&	https://cloudify.co	&	Operate	&	Orchestration	&	1	&	8	&	9	\\
Flyte	&	https://flyte.org/	&	Operate	&	Orchestration	&	1	&	16	&	17	\\
Aparavi	&	https://www.aparavi.com/	&	Plan	&	Data Management	&	1	&	0	&	1	\\
Databricks	&	https://databricks.com/	&	Plan	&	Data Management	&	1	&	0	&	1	\\
dbt	&	https://www.getdbt.com	&	Plan	&	Data Management	&	3	&	12	&	15	\\
Rivery	&	https://rivery.io/product/	&	Plan	&	Data Management	&	0	&	2	&	2	\\
Alectio	&	https://alectio.com	&	Plan	&	Data Management	&	0	&	1	&	1	\\
Apache NiFi	&	https://nifi.apache.org	&	Plan	&	Pipeline Design	&	1	&	0	&	1	\\
Kedro	&	https://kedro.org	&	Plan	&	Pipeline Design	&	4	&	13	&	17	\\
Luigi	&	https://github.com/spotify/luigi	&	Plan	&	Pipeline Design	&	1	&	5	&	6	\\
Orchest	&	https://www.orchest.io	&	Plan	&	Pipeline Design	&	0	&	2	&	2	\\
Ansible	&	https://www.ansible.com	&	Release/Deploy	&	Configuration Management	&	4	&	3	&	7	\\
Chef	&	https://www.chef.io	&	Release/Deploy	&	Configuration Management	&	3	&	2	&	5	\\
Baseten	&	https://www.baseten.co	&	Release/Deploy	&	Deployment and Serving	&	0	&	1	&	1	\\
BentoML	&	https://bentoml.com/	&	Release/Deploy	&	Deployment and Serving	&	6	&	11	&	17	\\
Datatron	&	https://www.datatron.com/	&	Release/Deploy	&	Deployment and Serving	&	3	&	11	&	14	\\
OpenVino	&	https://docs.openvino.ai/latest/home.html	&	Release/Deploy	&	Deployment and Serving	&	1	&	2	&	3	\\
Seldon Core	&	https://www.seldon.io/solutions/open-source-projects/core	&	Release/Deploy	&	Deployment and Serving	&	1	&	2	&	3	\\
Cortex	&	https://www.cortex.dev	&	Release/Deploy	&	Deployment and Serving	&	5	&	21	&	26	\\
Terraform	&	https://www.terraform.io	&	Release/Deploy	&	Infrastructure Provisioning	&	3	&	1	&	4	\\
Selenium	&	https://selenium-python.readthedocs.io	&	Test	&	Browser	&	1	&	2	&	3	\\
DeepChecks	&	https://deepchecks.com	&	Test	&	ML Model	&	0	&	1	&	1	\\
\hline
\multicolumn{5}{l}{The value ``\#'' in the last 3 columns indicates  the number of mentions.}\\
\end{tabular}
}
\end{table*}
\end{document}